\documentclass[reprint,prl,superscriptaddress,showpacs,nofootinbib,%
tightenlines ]{revtex4-1}
\usepackage{epsfig}
\usepackage{amssymb,amsmath}

\newcommand{\ben}{\begin{displaymath}}
\newcommand{\een}{\end{displaymath}}
\newcommand{\be}{\begin{equation}}
\newcommand{\ee}{\end{equation}}
\newcommand{\bea}{\begin{eqnarray}}
\newcommand{\eea}{\end{eqnarray}}
\begin{document}
\title{Triviality of quantum electrodynamics revisited}
\author{D.~Djukanovic}
\affiliation{Helmholtz Institute Mainz, University of Mainz, D-55099 Mainz, Germany}
\author{J.~Gegelia}
\affiliation{Institute for Advanced Simulation, Institut f\"ur Kernphysik
   and J\"ulich Center for Hadron Physics, Forschungszentrum J\"ulich, D-52425 J\"ulich,
Germany}
\affiliation{Tbilisi State  University,  0186 Tbilisi,
 Georgia}
\author{Ulf-G.~Mei\ss ner}
 \affiliation{Helmholtz Institut f\"ur Strahlen- und Kernphysik and Bethe
   Center for Theoretical Physics, Universit\"at Bonn, D-53115 Bonn, Germany}
 \affiliation{Institute for Advanced Simulation, Institut f\"ur Kernphysik
   and J\"ulich Center for Hadron Physics, Forschungszentrum J\"ulich, D-52425 J\"ulich,
Germany}

\date{31 May, 2017}

\begin{abstract}
Quantum electrodynamics is considered to be a trivial theory.
This is  based on a number of evidences, both numerical and analytical.
One of the strong indications for triviality of QED is the existence of the
Landau pole for the running coupling. We show that by treating QED as the leading
order approximation of an effective field theory and including the next-to-leading order
corrections, the Landau pole is removed.
Therefore, we conclude that the conjecture, that for reasons of self-consistency, 
QED needs to be \emph{trivial} is a mere artefact of the leading order
approximation to the corresponding effective field theory.

\end{abstract}


\maketitle

The concept of triviality in quantum field theories originates from papers
by Landau and collaborators studying the asymptotic behaviour of the photon
propagator in quantum electrodynamics (QED) \cite{Landau1,Landau2} (for a
review see e.g. Ref.~\cite{Callaway:1988ya}). Resumming the leading logarithmic
contributions they found that the photon propagator has a pole at large
momentum transfer. If this pole persists also in non-perturbative calculations
then to avoid the apparent inconsistency  QED has to be a non-interacting, i.e.
\emph{trivial}, theory. In calculations applying a finite cutoff this problem
manifests itself as a singularity in the bare coupling for a finite value of
the cutoff. It is therefore impossible to remove the cutoff unless the
renormalized coupling vanishes.

The standard model, in a  modern point of view, is a leading order
approximation to an effective field theory (EFT)  \cite{Weinberg:mt}.  While
the effective Lagrangian contains an infinite number of local interactions
consistent with the underlying symmetries, at  low energies the contributions
in physical quantities of the interactions with coupling constants of negative mass
dimensions (i.e. non-renormalizable interactions in the traditional sense) are
suppressed by powers of the energy divided by a large scale characterising
those degrees of freedom which are not explicitly taken into account in the
effective theory. In the framework of EFT the solution of the leading order
Wilson renormalization group equations might be obstructed at very large
cutoffs, however, this should not be a hard problem because of irrelevant interactions
or omitted fields being important at short distances \cite{Weinberg:mt}.
Therefore inconsistencies in the renormalization group analysis of 
renormalizable quantum field theories, like QED or $\phi^4$ theory of
self-interacting scalars, might be absent in the corresponding EFT framework.  

In this letter we address the consequences of treating QED as a leading order
approximation of an EFT for the problem of \emph{triviality}. To that end we
analyse the contributions of the next-to-leading order interaction, i.e. dimension
five operator, the well-known Pauli term. We start with the most general U(1)
locally gauge invariant  effective Lagrangian of the electron (and the
positron) field $\psi$ interacting with the electromagnetic field $A_\mu$
\begin{align}
{\cal L}  &=   - \frac{1}{4}\, F^{\mu\nu} F_{\mu\nu} +  \bar \psi \left( i D
\hspace{-.7em}/\hspace{.1em}-m\right) \psi  \nonumber\\ &+ \frac{i\kappa}{2} \,
\bar\psi (\gamma^\mu\gamma^\nu-\gamma^\nu\gamma^\mu) \psi \, F_{\mu\nu} + {\cal
L}_{\rm ho}\,, \label{lagrangianQEDEFT}
\end{align}
where $m$ is the (bare) electron mass, $e$ is the (bare) electromagnetic charge,
$F_{\mu\nu}=\partial_\mu A_\nu-\partial_\nu A_\mu$, $D_\mu= \partial_\mu - i e
A_\mu $ and ${\cal L}_{\rm ho}$ contains an infinite number of terms with
operators of order six and higher. We assume that the contributions of these
terms in the photon self-energy are suppressed compared to those of the
Pauli term. We remark  here that the standard QED Lagrangian given by the 
first line of Eq.~(\ref{lagrangianQEDEFT}) describes the experimental data very well. 
From the modern point of view this is because the contributions of terms in the 
second line are beyond the current accuracy of the data. In particular, 
the anomalous magnetic moment   of the electron in standard QED gets 
contributions only from loop diagrams and its calculated value agrees with the 
experiment very well suggesting that the Pauli term is suppressed by a scale larger 
than $4\times 10^7$ GeV \cite{Weinberg:mt}.

\medskip

The scale-dependent renormalized running coupling $e_R(q^2)$
can be defined by the following relation\footnote{We carried out all calculations 
in Landau gauge, however, the results are gauge independent.}
\begin{equation}
 D^{\mu\nu}(q) \, e^2=-\frac{1}{q^2}\left(g^{\mu\nu}-\frac{q^\mu q^\nu}{q^2}\right)\,  e_R^2(q^2), 
\label{defrcoupling}
\end{equation}
where $D^{\mu\nu}(q)$ is the dressed propagator of the bare photon field.
Calculating the dressed propagator at one-loop order we obtain for $-q^2\gg m^2$: 
\begin{eqnarray}
e_R^2(q^2)&=&\frac{e_r^2}{1-\frac{e_r^2
   +2 \kappa ^2 q^2}{12
   \pi ^2}\,\ln \frac{-q^2}{m^2} +c_R \, q^2 }\,.
\label{dPPR}
\end{eqnarray}
Here, $e_r$ is the renormalized coupling at $q^2=-m^2$ for $c_R=0$, where $c_R$
is a renormalized coupling constant of the higher-order Lagrangian ${\cal
L}_{\rm ho}$. It is suppressed by two orders of some large scale.  For
$\kappa=c_R=0$ the running coupling has the well-known pole singularity at (the
Landau pole) 
\begin{equation}
q^2_L= - m^2 \exp\left[12 \pi^2/e_r^2\right]~.
\label{Lpole}
\end{equation}
While this pole appears at extremely high energies, reducing its
practical importance to nil, it is still a problem if present in the full
theory \cite{Callaway:1988ya}. 
For reasonable values of $\kappa \gg  1/\sqrt{-q^2_L}$ the Landau pole is
absent remedying the inconsistency at the level of an EFT.
\begin{figure}[t]
\begin{center}
\epsfig{file=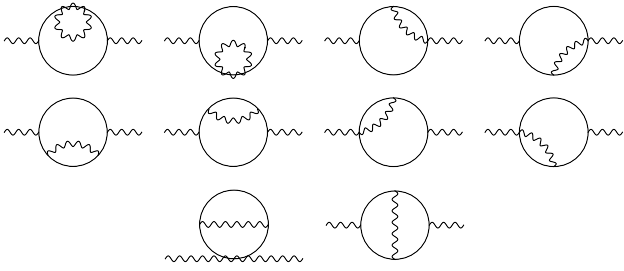,scale=0.5}
\caption{Two-loop diagrams contributing in the photon self-energy. The solid and wiggly lines correspond to the electron and photon propagators, respectively.}
\label{Two-Loop:fig}
\end{center}
\end{figure}
\begin{figure}[t]
\begin{center}
\epsfig{file=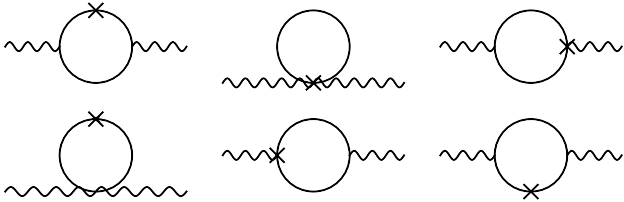,scale=0.5}
\caption{One loop counter term diagrams contributing in the photon self-energy at two-loop order. The solid and wiggly lines correspond to the electron and photon propagators, respectively. 
The crosses denote the counter terms of the one-loop order. Diagrams with ghost propagators are not shown. }
\label{One-LoopCT:fig}
\end{center}
\end{figure}
\medskip

In renormalizable theories only logarithmic divergences contribute
to the renormalization of coupling constants and therefore there is a direct
correspondence between the Gell-Mann-Low \cite{GellMann:1954fq} and the
Wilsonian renormalization group approaches \cite{Wilson:1973jj}. As a result, 
the presence of the Landau pole in the expression of the running renormalized 
coupling authomatically leads to  the pole in the bare coupling as a function 
of the cutoff parameter.  However, in an
EFT with non-renormalizable interactions the direct link between the two 
renormalization group equations is lost and therefore the Wilsonian 
renormalization group approach requires additional study of 
the cutoff dependence. 
We investigate the cutoff dependence of the bare electromagnetic coupling in our
model by applying the higher derivative regularization
\cite{Faddeev:1980be,Djukanovic:2004px}  which preserves the local U(1) gauge invariance. Notice that dimensional  regularization is not suitable here as it discards the power-law divergences.
In addition to the fields in conventional QED, we introduce scalar (i.e.
commuting) ghost fields $\bar\xi$ and $\xi$ which regulate the one-loop counter
term diagrams contributing to the photon self-energy at two-loop order.  The
effective Lagrangian generating one and two-loop diagrams, contributing to our
calculation of the photon self-energy up to two-loop $e^2\kappa^2$ order, which
are all finite for finite $\Lambda$, is given by:
\begin{align}
{\cal L}_{\rm HDR} & =   - \frac{1}{4}\, F^{\mu\nu}\left(1+\frac{\partial^2}{\Lambda^2}\right)^2
F_{\mu\nu} 
\nonumber\\
&+ \frac{1}{2} \, \bar \psi \left( i D \hspace{-.7em}/\hspace{.1em}-m\right)\left(1+\frac{D^2}{\Lambda^2}\right)^3 \psi + {\rm h.c.} 
\nonumber\\
&+ \frac{1}{2} \, \bar \xi \left( i D \hspace{-.7em}/\hspace{.1em}-m\right)\left(1+\frac{D^2}{\Lambda^2}\right)^3 \xi + {\rm h.c.} 
\nonumber\\
&+ \frac{i\kappa}{2} \, \bar\psi (\gamma^\mu\gamma^\nu-\gamma^\nu\gamma^\mu) \psi \, F_{\mu\nu} + {\cal L}_{\rm ho}.
\label{lagrangianQEDEFT_HDR}
\end{align}

The bare electromagnetic coupling as a function of the cutoff satisfies a
renormalization group equation which up to the level of accuracy of our
calculation  has the form:
\begin{equation}
\frac{d \alpha(\Lambda)}{d\ln\Lambda} = A_1\,\alpha^2(\Lambda) + \kappa^2 \alpha^2(\Lambda) \,\Lambda^2(2 A_2 + A_3 +2 A_3 \ln\Lambda/m),
\label{REq}
\end{equation}
where $\alpha(\Lambda)=e^2/(4\pi)$, the coefficient $A_1$ is given by one-loop
diagrams and $A_2$ and $A_3$ are extracted from two-loop calculations. Notice
that there are no power-law divergences at one-loop order and all terms
suppressed by powers of $m/\Lambda$ have been dropped in our calculations as
they are negligible for large values of $\Lambda$.

\medskip

The solution to Eq.~(\ref{REq}) is given by
\begin{widetext}
\begin{align}
\alpha(\Lambda) & =  \frac{\alpha _0}{1-\alpha _0 \ln \frac{\Lambda }{m}  \left(A_1+ A_3 \kappa ^2
   \Lambda ^2\right) 
   + \alpha _0 \ln \frac{\Lambda _0}{m}
   \left(A_1+A_3 \kappa ^2 \Lambda _0^2\right) - \alpha _0 A_2 \kappa ^2
   (\Lambda ^2-\Lambda _0^2) },
\label{runnc}
\end{align}
\end{widetext}
where $\alpha_0=\alpha(\Lambda_0)$  is the bare coupling at some fixed cutoff 
$m < \Lambda_0< \Lambda$.

For $\Lambda\gg \Lambda_0$ we have
\begin{eqnarray}
\alpha(\Lambda) = \frac{\alpha _0}{ 1-\alpha _0 \ln \frac{\Lambda }{m}  \left(A_1+ A_3 \kappa ^2
   \Lambda ^2\right)  
   - \alpha _0 A_2 \kappa ^2
   \Lambda ^2 
    }\,.
\label{runncLggL0}
\end{eqnarray}
For $\kappa=0$ and positive $A_1$ the expression in Eq.~(\ref{runncLggL0}) has a pole at 
\begin{equation}
\Lambda_{\rm P}=m \, \exp\Bigl[\frac{1}{A_1\alpha_0}\Bigr].
\label{lpole}
\end{equation}
This pole, if remaining in the non-perturbative full expressions of the bare
coupling, prevents the $\Lambda\to \infty$ limit unless $\alpha_0\equiv 0$, thus
leaving us with a non-interacting theory.  In renormalizable theories where
only logarithmic divergences contribute in the renormalization of coupling
constants, there is a close correspondence between the Gell-Mann-Low and the
Wilsonian renormalization group approaches manifesting itself in a direct
relation of Eqs.~(\ref{Lpole}) and (\ref{runncLggL0}) valid for  standard
QED.    

Using FeynCalc \cite{Mertig:1990an,Shtabovenko:2016sxi} and Form
\cite{Vermaseren:2000nd} and applying the method of dimensional counting of
Ref.~\cite{Gegelia:1994zz} we have calculated the logarithmically divergent
contributions to the photon self-energy generated by one-loop diagrams, and
the quadratic divergences generated by the two-loop diagrams, shown in
Fig.~\ref{Two-Loop:fig},\footnote{Notice here that due to the higher derivative
regularization there are interaction vertices not present in standard QED.}
and by the corresponding counter term diagrams, shown in
Fig.~\ref{One-LoopCT:fig}. Our results read:\footnote{We have used HypExp2
\cite{Huber:2007dx} to expand the hypergeometric functions in a Laurent series
in $\epsilon$.}
\begin{align}
 A_1 & =  \frac{2}{3 \pi} \simeq 0.212,  \ \ A_3=-\frac{7}{40 \, \pi ^3}\simeq -0.0056,\nonumber\\
 A_2 &=
-\frac{5491889}{201600 \, \pi ^3} -\frac{2194315}{104976 \, \pi } +\frac{8593 \ln 3}{5832 \sqrt{3} \, \pi ^2}  \nonumber \\
& +  \frac{1508572 \,\psi ^{(1)}\left(\frac{1}{3}\right)-303413 \, \psi
   ^{(1)}\left(\frac{1}{6}\right)}{17496 \, \pi ^3} \nonumber\\   
& + \frac{8593 i }{972 \sqrt{3} \, \pi ^3}  \biggl[  2 \, \text{Li}_2\left(-\frac{i}{\sqrt{3}}\right)-2\,
   \text{Li}_2\left(\frac{i}{\sqrt{3}}\right) \nonumber\\
 &  -\text{Li}_2\left(\frac{1}{2}-\frac{i}{2 \sqrt{3}} \right)+\text{Li}_2\left(\frac{1}{2}+\frac{i}{2 \sqrt{3}}\right) \biggr]
\simeq 0.0040 \nonumber\, ,
  \label{coeffs}
\end{align}
where $\text{Li}_2$  and $\psi^{(1)}$ are the dilogarithm and trigamma functions, 
respectively. For the above values of
$A_1$, $A_2$ and $A_3$,  and for \emph{natural} values of $\kappa \gg
1/\Lambda_{\rm P}$, in the denominator of Eq.~(\ref{runncLggL0}) the $A_3$ term 
is larger than the $A_1$ and $A_2$ terms and the negative sign of $A_3$ 
guarantees that $\alpha(\Lambda)$ has no pole.

\medskip

To conclude, we have shown that the problem of \emph{triviality} in QED 
can be attributed to QED being a leading order approximation of an effective
field theory. We have shown that already at next-to-leading order, i.e. adding
the Pauli term to QED, the Landau pole disappears thereby obviating the need for
QED to be a \emph{trivial} theory.

\medskip

\acknowledgments

This work was supported in part by DFG and NSFC through funds provided to the
Sino-German CRC 110 ``Symmetries and the Emergence of Structure in QCD" (NSFC
Grant No.~11621131001, DFG Grant No.~TRR110), by the Georgian Shota Rustaveli National
Science Foundation (grant FR/417/6-100/14) and by the CAS President's International
Fellowship Initiative (PIFI) (Grant No.~2017VMA0025).



\end{document}